\newcommand{\be}{\begin{equation}}
\newcommand{\ee}{\end{equation}}
\newcommand{\ber}{\begin{eqnarray}}
\newcommand{\eer}{\end{eqnarray}}
\newcommand{\gsim}{\raisebox{-0.7ex}{$\stackrel{\textstyle >}{\sim}$ }}
\newcommand{\lsim}{\raisebox{-0.7ex}{$\stackrel{\textstyle <}{\sim}$ }}
\def\fm3{fm$^{-3}$}

\documentstyle[prl,aps,epsfig]{revtex}

\draft
\begin{document}
\title{First Order Phase Transitions in Neutron Star Matter:\\
Droplets and Coherent Neutrino Scattering}
\author {Sanjay Reddy$^1$ , George Bertsch$^1$  and 
Madappa Prakash$^2$ }
\address{$^1$Institute For Nuclear Theory, University of Washington, Seattle,
WA 98195. \\ $^2$Dept. Physics \& Astronomy, SUNY at Stony Brook, Stony Brook,
NY 11794. }
\date{\today}
\maketitle
\begin{abstract}
A first order phase transition at high baryon density implies that a mixed
phase can occupy a significant region of the interior of a neutron star. In
this article we investigate the effect of a droplet phase on neutrino
transport inside the core. Two specific scenarios of the phase transition are
examined, one having a kaon condensate and the other having quark matter in the
high density phase. The coherent scattering of neutrinos off the droplets
greatly increases the neutrino opacity of the mixed phase. We comment on how
the existence of such a phase will affect a supernova neutrino signal.
\end{abstract}
\pacs{PACS numbers(s): 13.15.+g, 26.60.+c, 97.60.Jd} 
Neutrino interactions play a central role in the early evolution of neutron
stars formed during a type II supernova. The enormous binding energy ($\sim
10^{53}$ ergs) gained during implosion of the inner core of a massive star is
released in the form of neutrinos.  Temporal and spectral characteristics of
the neutrino emission from the newly born neutron star, also called a
protoneutron star, influences the explosion mechanism, the r-process
nucleosynthesis, and more importantly, is an observable feature of a supernova
explosion. The handful of neutrino events observed from SN87A are testimony 
to this fact.

In this article we study how a first order phase transition at high density
will influence the neutrino opacity. We consider two models of high density
matter wherein a first order phase transition occurs for densities of relevance
for neutron stars: first order kaon condensation and the quark-hadron
transition.  The mixed phase consists of a high baryon density, negatively
charged, kaon condensed matter or quark matter coexisting with lower density,
positively charged, baryonic matter. We study the region of the mixed phase 
where the structure is expected to be droplets of kaonic matter or quark
matter embedded in a lower baryon density nucleonic matter. Our primary
motivation for this study is to determine the neutrino opacity of this material
since this is directly related to an observable - the supernova neutrino
luminosity curve.

Our main finding is that the neutrino mean free path in a mixed phase
containing either kaon condensed matter or quark matter is greatly reduced
compared to uniform baryonic matter at the same density. This is a consequence
of the coherent scattering of neutrinos from the matter in the droplets. The
droplets carry a large weak charge and scattering of long-wavelength neutrinos
increases as its square.  The reduction of the mean free path compared to that
in pure neutron matter could be as large as one to two orders of
magnitude. Thus, despite our poor knowledge of neutrino opacities in dense
nucleonic matter, we argue that the presence of a first order phase transition
is likely to have a dramatic and discernible effect on the temporal
characteristics of supernova neutrino light curve.

{\it Neutrino-Droplet Scattering}: Prior to discussing in detail the specific
models in which a droplet phase is energetically favored, we describe
neutrino-droplet scattering.  Neutrinos scatter off the droplets as they carry
a net excess of weak-charge. For the typical droplet sizes which range from
$5-15$ fm we may infer that the scattering will be fairly coherent for momentum
transfers $ \lsim 40$ MeV.

First, we consider neutrino scattering from isolated droplets. The Lagrangian
that describes the neutral current coupling of neutrinos to the droplet is
given by
\be
{\mathcal{L}}_W = \frac{G_F}{2\sqrt{2}} ~\bar{\nu}\gamma_\mu(1-\gamma_5) \nu 
~J^{\mu}_D \,.
\label{nuD}
\ee
where $J^{\mu}_D$ is the neutral current carried by the droplet and $G_F=1.166
\times 10^{-5}$ GeV$^{-2}$ is the Fermi weak coupling constant. For 
non-relativistic droplets $J^{\mu}_D = \rho_W(x)~\delta^{\mu 0}$ has only a
time like component and $\rho_W(x)$ is the excess weak charge density in the
droplet. The total weak charge enclosed in a droplet of radius $r_d$ is given
by $N_W=\int_0^{r_d} d^3x ~\rho_W(x)$ and the form factor is
$F(q)=(1/N_W)\int_0^{r_d} d^3x ~\rho_W(x)~ \sin{qx}/qx$. The differential cross
section for neutrinos scattering from an isolated droplet is then given by
\be 
\frac{d\sigma}{d\cos{\theta}}= \frac{E_\nu^2}{16\pi} G_F^2 N^2_W(1+\cos{\theta}) F^2(q) \,. 
\label{diff}
\ee
In the above equation $E_\nu$ is the neutrino energy and $\theta$ is the
scattering angle. Since the droplets are massive we consider only elastic
scattering in which case the magnitude of the momentum transfer is given by
$q=\sqrt{2}E_\nu(1-\cos{\theta})$.

{\it Neutrino transport in the mixed phase:} We must also embed the droplets
into a description of the medium to evaluate the neutrino transport parameters.
The droplet radius $r_d$ and the inter-droplet spacing are determined by the
interplay of surface and Coulomb energies and depends on the specific details
of the high density model. We will calculate the droplets in the Wigner-Seitz
cell assuming that the medium has a droplet density $N_D$. The cell radius
$R_W=(3/4\pi N_D)^{1/3}$. Except for one aspect, we will neglect the coherent
scattering from more than one droplet. If the droplets form a lattice, Bragg
scattering will dominate and our description would not be valid. But for a low
density and a liquid phase the interference from multiple droplets affects only
the scattering at long wavelengths. If the ambient temperature is not small
compared to the melting temperature the droplet phase will be a liquid and for
neutrino energy in the range $E_\nu \gsim (1/R_W) $ interference effects
arising from scattering off different droplets is small. However, multiple
droplet scattering cannot be neglected for $E_\nu \lsim 1/R_W$. The effects of
other droplets is to cancel the scattering in the forward direction, because
the interference is destructive except at exactly zero degrees, where it
produces a change in the index of refraction of the medium. We account for this
by subtracting from the weak charge density $\rho_W$ a uniform density which
has the same total weak charge $N_W$ as the matter in the Wigner-Seitz
cell. The effect on the form factor is to modify it as follows
\ber 
F(q) \rightarrow \tilde{F}(q) = F(q) - 3~
\frac{\sin{qR_{W}} - (qR_{W})\cos{qR_{W}}}{(q R_{W})^3}  \,.
\label{formc}
\eer
In the medium, the appropriate differential cross section is defined per unit
volume rather than per droplet. The neutrino-droplet differential cross section
per unit volume follows from the preceding discussion and is given by
\be 
\frac{1}{V}\frac{d\sigma}{d\cos{\theta}}= N_D~\frac{E_\nu^2}{16\pi} G_F^2 N^2_W(1+\cos{\theta}) \tilde{F}^2(q) \,. 
\label{diff1}
\ee
Note that even for small droplet density $N_D$ the factor $N_W^2$ acts to
enhance the droplet scattering. To quantify the importance of droplets as a
source of opacity we compare with the standard scenario where matter is uniform
and composed of neutrons. In this case the dominant source of opacity is due to
scattering from thermal fluctuations and the cross section per unit volume is
given by
\ber
\frac{d\sigma}{V d\cos{\theta}}&=&\frac{G_F^2}{8\pi}(c_V^2(1+\cos{\theta})
+(3-\cos{\theta})c_A^2) ~ E_{\nu}^2 \nonumber \\
&\times& \frac{3}{2}~ n_n ~\left[\frac{k_BT}{E_{fn}}\right]\,,
\label{diff2}
\eer
where $c_V,c_A$ are the vector and axial coupling constants of the neutron,
$n_n$ is neutron number density, $E_{fn}$ is the neutron Fermi energy
and $T$ is the matter temperature \cite{S,RPL}.

The transport cross sections that are employed in studying the diffusive
transport of neutrinos in the core are weighted by the angular factor
$(1-\cos{\theta})$. The transport mean free path $\lambda(E_\nu)$ for given
neutrino energy $E_\nu$ is given by
\ber 
\frac{1}{\lambda(E_\nu)}=\frac{\sigma_T(E_\nu)}{V}=\int d\cos{\theta}~ (1-\cos{\theta})
\left[\frac{1}{V} \frac{d\sigma}{d\cos{\theta}}\right] \,. 
\label{fint}
\eer
It is thus the task of the models to provide the weak charge and form factors
of the droplets so that we may calculate the contribution of neutrino-droplet
scattering to the opacity of the mixed phase.

 {\it First-order Kaon condensation} Kaon condensation is one of the several
possible forms of exotica that could exist at high baryon density. Since the
pioneering work of Kaplan and Nelson \cite{KN} several authors (see
\cite{KC1,KC2} and references therein) have studied in detail the role of such a
condensate in neutron star structure and evolution. Kaon condensation
significantly softens the high density equation of state (EOS) thereby lowering
the maximum mass of neutron stars. Glendenning and Schaffner \cite{GS}
revisited the problem and studied kaon condensation in a relativistic mean
field model, wherein the kaon is minimally coupled to $\sigma$, $\omega$ and
$\rho$ mesons. Within the purview of this model they were able to show that
kaon condensation occurs as a first order phase transition.

A relativistic field theoretical approach is used to describe the baryons.  We
follow the model employed by Glendenning and Schaffner \cite{GS} wherein the
Lagrangian density is given by
\begin{eqnarray*}
L &=& \sum_{B} \overline{B}(i\gamma^{\mu}\partial_{\mu} 
- g_{\omega B} \gamma^{0}\omega_0
-g_{\rho B}\gamma^{0}{\bf{b}}_{0}\cdot{\bf t} \\
&-& M_B+g_{\sigma B}\sigma)B \\
&+& \frac{1}{2}m_{\omega}^2\omega^2_{0}
+ \frac{1}{2}m_{\rho}^2 b_{0}^2
-\frac{1}{2}m_{\sigma}^2\sigma^2-U(\sigma) \,.
\end{eqnarray*}
Here, $B$ are the Dirac field operators for baryons and ${\bf t}$ is the
isospin matrix. In the mean field approximation only the zeroth components of
the sigma, omega and rho meson fields are retained and are denoted by $\sigma,
\omega_0$ and $b_0$ respectively. $U(\sigma)$ represents the scalar
self-interactions and is of the form $U(\sigma) = (b/3)M_n(g_{\sigma
N}\sigma)^3 +(c/4)(g_{\sigma N}\sigma)^4$.  Kaons are coupled minimally to the
meson fields appearing in the baryonic part of the Lagrangian. The Lagrange
density for the kaon field is given by
\begin{eqnarray*}
L_K={\cal{D}}_{\mu}^*K^*{\cal{D}}^{\mu}K 
- (m_K - g_{\sigma K} \sigma)^2 K^*K\,, 
\end{eqnarray*}
where the covariant derivative 
\begin{eqnarray*}
{\mathcal{D}}_\mu
=(\partial_{\mu} + i(g_{\omega K} \omega_\mu + g_{\rho K} b_\mu)) \,.
\end{eqnarray*}
Droplets are constructed in the Thomas-Fermi or local density
approximation. The equation of motion for the kaons is given by
\ber
\left[{\mathcal{D}}_{\mu}{\mathcal{D}}^{\mu} + (m_k^*)^2\right] K^-=0 \,,
\label{eom}
\eer
where $m_K^*=m_K-g_{\sigma K} \sigma$ is the kaon effective mass in the
medium. Eq. (\ref{eom}) is solved by writing explicitly the spatial and
temporal dependence of the kaon field as $K^-= f_\pi \theta(r) \exp(i \mu_K t)
/\sqrt{2}$ where $f_\pi=93$ MeV is the pion decay constant. The spatial part
has only a radial dependence, and the time dependence is fixed by the chemical
potential \cite{BYM}. In principle we should solve a Klein-Gordon equation for
the other meson fields as well. However, the other mesons are significantly
more massive and their fields do not vary over distances of the order of their
Compton wavelengths. In the vicinity of the critical density the effective kaon
mass is approximately given by $m^*_K \sim \mu_e \sim$ 200 MeV and is small
compared to $\sigma$, $\omega$ or $\rho$ meson masses. In this approximation
the non-strange meson field equations are
\ber
m^2_\sigma \sigma &=& -\frac{dU}{d\sigma} + g_{\sigma B}(\rho^s_n
+\rho^s_p)
+ g_{\sigma k} f_\pi^2 \theta^2 m_k^* \, ,\\
m_\omega^2 \omega_o &=& g_\omega (\rho_n+\rho_p)
   - g_{\omega K}^2 f^2_\pi \theta^2 (\mu_K+X) \, ,\\
m_\rho b_o &=& \frac{1}{2} g_\rho (\rho_p-\rho_n)
   - g_{\rho K} f^2_\pi \theta^2 (\mu_K+X) \, .
\eer
where $\rho_n,\rho_p,\rho^s_n$ and $\rho^s_p$ are the baryon number and scalar
densities of the neutrons and protons respectively and the vector potential for
the kaons is given by $X=g_{\omega k} \omega_o + g_{\rho k} b_o $.

The classical kaon field equation is obtained by substituting the specific form
given above in Eq.(\ref{eom}) and is given by
\be
-\nabla^2 \theta(r) + [(\mu_K+X)^2 - (m_k-g_{\sigma} \sigma_o)^2] \theta(r)=0 \, .
\label{dife}
\ee
The above differential equation is solved with appropriate boundary conditions
to obtain the droplet configuration. The baryon and electron chemical
potentials are close to those in the mixed phase, but must be fine tuned to
give a solution to Eq.(\ref{dife}) that satisfies the boundary conditions at
both $r=0$ and $r=R_W$. The energy is computed in the Wigner-Seitz
approximation and the droplet radius that minimizes the energy density is
determined. The energy density contains bulk, surface and Coulomb
contributions. The baryonic contribution is
\be 
\epsilon_B(r) = \epsilon_{kin} + \frac{1}{2} m_\sigma^2 \sigma_o^2 
+ \frac{1}{2} m_\omega^2 \omega_o^2 
+ \frac{1}{2} m_\rho^2 b_o^2 + U(\sigma_o) \,,
\ee
and the kaon contribution is given by
\be
\epsilon_k(r)=f_\pi^2 \theta^2(r) (\mu_K + X)^2          \,.
\ee
Note that the surface contribution is contained in the above expression and
that terms proportional to the gradients have been eliminated by using the
equation of motion for the kaons.

The shape of the cell depends on the type of lattice that is favored; for BCC
or FCC the Wigner-Seitz cell is a regular polyhedron and can be approximated by
a sphere.  The Wigner-Seitz radius, denoted by $R_W$, is determined by size of
the cell that encloses zero total electric charge.  Since the electric field
vanished at $r=R_W$, the Coulomb energy is given by
\be
E_c = \frac{\alpha}{2} \int_o^{R_W} dr \frac{Q^2_{e}(r)}{r^2} \, ,
\ee
where $\alpha=1/137$ and $Q_e(r)$ is the enclosed electric charge a distance
$r$ from the center.

We find that the Debye screening lengths $(1/\lambda^2_D)_{i}=4\pi \alpha
(\partial n_i/\partial \mu_i)_{n_j,j \ne i}$ of electrons, protons and kaons
are of the same order as that the typical droplet radius.  Thus the charged
particle distributions may change and screen the droplet charge
\cite{HPS}. We allow for this by adding the Coulomb term to the single
particle energies and employing a iterative procedure to relax the charged
particle distributions to their equilibrium value. For droplets radii $r_d \sim
\lambda_D$ the charge particle distributions in the droplet changes at the
10-30 percent level depending on their individual Debye screening
lengths. However, for $r_d/\lambda_D \gg 1$ the charge particle distributions
change drastically and all the charge resides in the surface region. For these
conditions the droplet phase in not energetically favored \cite{HPS}. Our
investigation indicates that for baryon density in the region $0.52-0.62$
fm$^{-3}$ the droplets survive, but with moderate modifications to the charged
particle profiles. Fig. 1 shows the particle density profiles at $n_B=0.54$
fm$^{-3}$.

\begin{figure}[h]
\centering
{\epsfig{figure=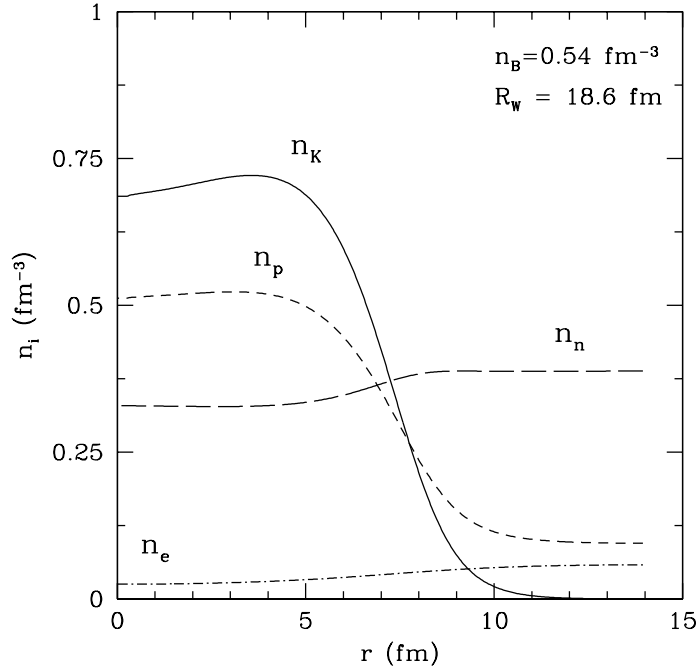,width=.55\textwidth}}
\caption[]{Droplet profile at baryon density $n_B=0.54$ fm$^{-3}$ . 
The solid curve shows the kaon number density, the short-dashed and long-dashed
curves show the proton and neutron number densities respectively. The kaon,
proton, neutron and electron number densities are shown as solid, dashed,
long-dashed and dot-dashed curves respectively.}
\label{fig1}
\end{figure}

The excess vector weak charge density of the droplet is
\ber
\rho_W(r) &=& (-1 + 2\sin^2{\theta_W}) n_K(r) \nonumber \\
&+& (1+4\sin^2{\theta_W}) (n_e(r) - n_e^o) \nonumber \\
 &+& (1 - 4\sin^2{\theta_W})(n_p(r) - n_p^o) 
- (n_n(r) - n_n^o) \,,
\eer
where $n_K,n_e,n_p$ and $n_n$ are the kaon, electron, proton and neutron number
densities respectively and $n_e^o,n_p^o$ and $n_n^o$ are corresponding
densities in the low density phase. Kaons dominate the weak charge density
since (i) the electron number is small, (ii) the protons carry negligible weak
charge, and (iii) the neutron number density inside and outside the droplet are
nearly equal. The neutral current neutrino-kaon coupling is not experimentally
measured and our estimates here are based on the constituent quark model (note
that the main source of uncertainty is the singlet contribution \cite{MW}, in
the large $N_C$ limit that the singlet current contributes at the 30 percent
level to the weak charge of the kaon). The total weak charge of the droplet
shown in Fig. 1 is $N_W \simeq 700$. The corresponding form factor $F(q)$ is
shown in the top panel of Fig. 2.

The transport differential scattering (Eq.(\ref{diff})) weighted with the
factor $(1-\cos{\theta})$ for $E_\nu= 50$ MeV and for a baryon density of
$n_B=0.54$ fm$^{-3}$ is shown in the bottom panel of Fig. 2. For comparison we
also show results for uniform neutron matter (computed by using
Eq.(\ref{diff2}) and for $T=10$ MeV) where thermal density fluctuations
dominate the opacity.
\begin{figure}[b]
\centering
{\epsfig{figure=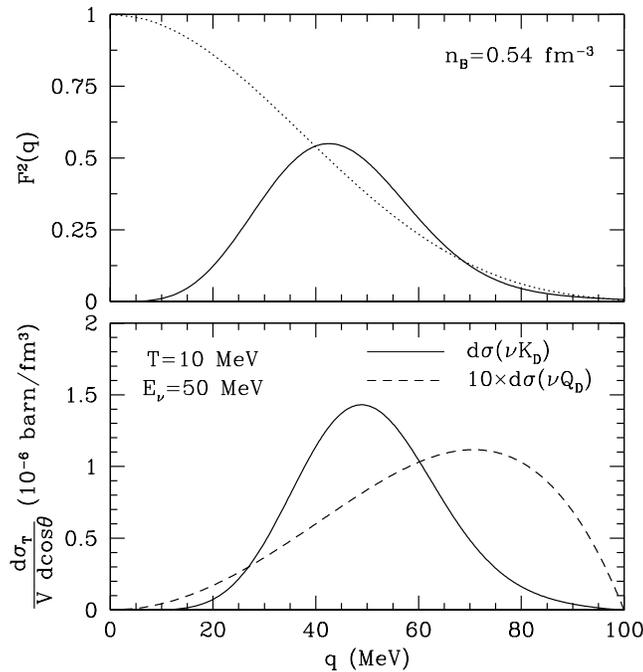,width=.5\textwidth}}
\caption{Top panel: The dashed curve shows the form factor for a single drop 
and the solid curve shows the results for an embedded droplet
(Eq.(\ref{formc})).  Lower panel: Transport differential cross section for
scattering off droplets (solid curve) and for scattering off thermal
fluctuations in pure neutron matter (dashed curve). Note that dashed curve has
been multiplied by a factor of 10.}
\end{figure}
The neutrino mean free path computed using Eq.(\ref{diff1}) is shown (solid
curve) in Fig. 3 for $n_B=0.54$ and for a temperature of $10$ MeV. The mean
free path in uniform neutron matter is also shown (dashed curve) for
comparison. As is evident from the figure, the mean free path in the mixed
phase is significantly reduced compared to pure neutron matter for typical
neutrino energy $E_\nu \sim \pi T$. The reduction is severe at moderate
energies, and is easily understood by noting that the form factors have
significant support in the region where $q\sim 40$ MeV. At lower $q$ the
inter-droplet correlations act to screen the weak charge of the droplet and at
higher energy is attenuated by the form factor. Although the transport
calculations would require the energy dependent mean free path it is useful to
define an average mean free path for energy transport $<\lambda_E>=\int
dE_\nu~f_\nu(E_\nu)~E_\nu^4
\lambda(E_\nu)/ (\int dE_\nu~f_\nu(E_\nu)~E_\nu^4)$ where $f_\nu(E_\nu)$ is the neutrino distribution function \cite{RPL}. At $T=10 MeV$we find that
$<\lambda_E>$ is reduced by a factor $\sim 20$ and at $T=30$ MeV the 
reduction is reduced by a factor $\sim 9$.

\begin{figure}[]
\centering
{\epsfig{figure=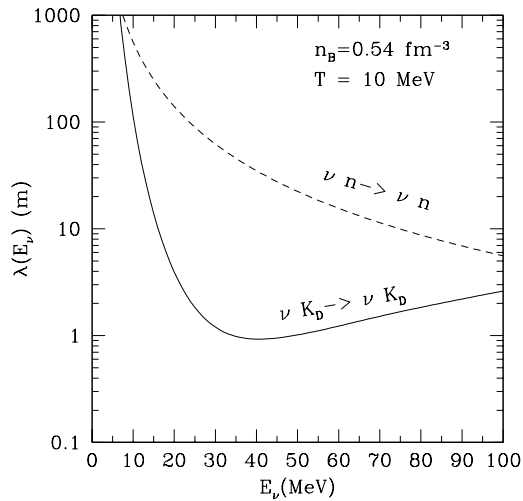,width=.4\textwidth}}
\caption{Neutrino mean free path in the droplet phase. Results for neutrino droplet scattering (solid curve) are compared with the uniform neutron matter 
case (dashed curve).}
\label{elam1}
\end{figure}

{\it First order quark-hadron transition} Several authors have studied the
possibility of a first order quark-hadron phase transition in a bag model
picture for densities of relevance to neutron star interiors \cite{JP,G,PCL}.
These studies found that a mixed phase was energetically favored. Detailed
investigations of the structured phase by Heiselberg {\it et al.}, wherein
surface and Coulomb contributions to the energy density were properly accounted
for, showed that the presence of a droplet phase was sensitive to the
value of the surface tension $\sigma$ between the quark and hadron phases
\cite{HPS} . For $\sigma \le 100$ MeV the droplet sizes favored were smaller 
than the typical Debye screening length $\lambda_D \sim 7$ fm and the droplet
phase was energetically favored.

The droplet radius is given by $r_d = \left[15 \sigma/(8\pi
(\rho_Q-\rho_N)^2)\right]^{1/3}$, where $\rho_Q$ and $\rho_N$ are the electric
charge densities in the quark and nucleon phases respectively. The charge
densities depend in general on details of the model and a typical scale for the
difference $\rho_Q-\rho_N \sim0.4$ e fm$^{-3}$. For $\sigma$ in the range
$10-100$ MeV this corresponds to droplet radii of $r_d=3-7$ fm.  For our
estimate of coherent scattering we will assume a droplet radius $r_d=5$ fm at a
baryon density $n_B=0.7$ fm$^{-3}$. In a minimal model the nucleonic phase is
described by a relativistic mean field theory and the quark phase as a free
Fermi gas with a bag constant \cite{PCL} $B=200$ MeV/fm$^3$ and a finite
strange quark mass $m_s=150$ MeV.  In this model, and for $n_B=0.7$ fm$^{-3}$,
the baryon density in the nucleon phase is $n_N = 0.68$ fm$^{-3}$ and $n_Q =
0.94$ fm$^{-3}$ in the quark phase. The volume fraction occupied by the quark
phase is given by $ f = (n_B-n_N)/(n_Q-n_N)
\sim 0.1$ and the Wigner-Seitz cell radius spacing $R_W= r_d/f^{1/3} \sim
11$ fm.

\begin{figure}[]
\centering
{\epsfig{figure=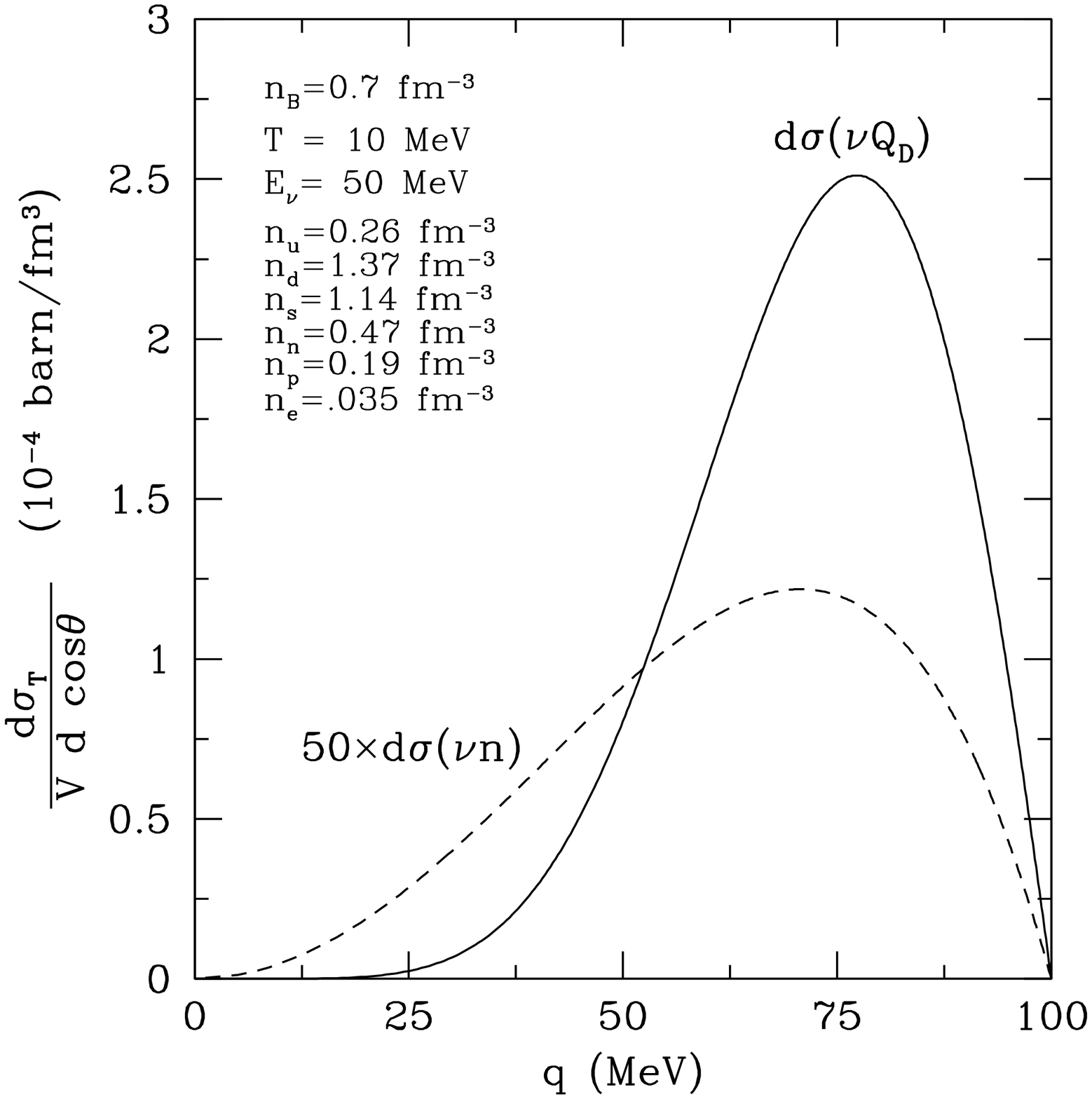,width=.48\textwidth}}\quad
{\epsfig{figure=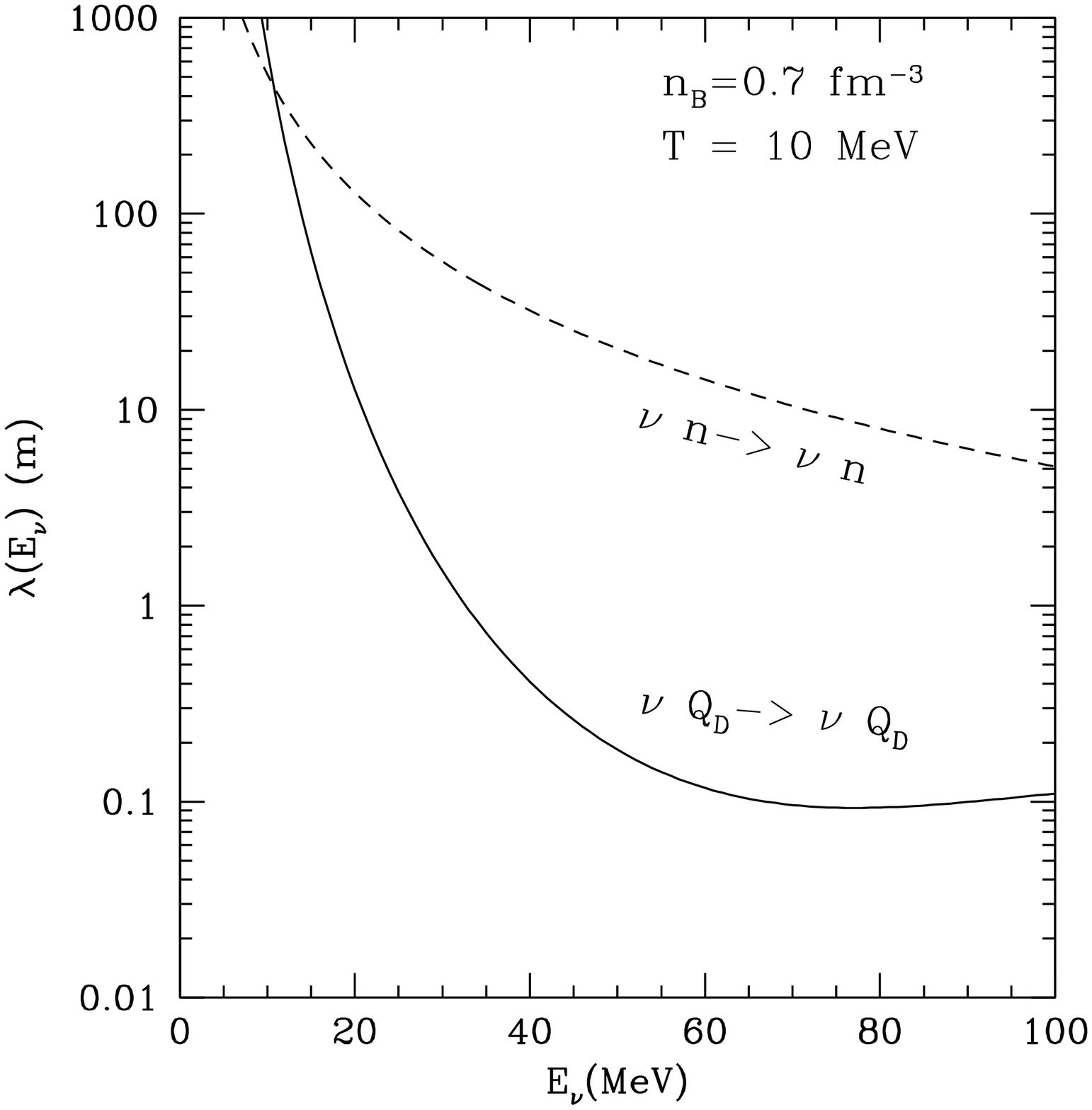,width=.48\textwidth}}
\caption{Results for 
neutrino-quark droplet scattering (solid curve) are compared with the uniform
neutron matter case (dashed curve). Left panel shows the results for the
differential scattering of neutrinos with energy $E_\nu=50$ MeV and the right
panel shows the neutrino mean free path at $n_B=0.7$ fm$^{-3}$ and $T=10$
MeV. The left panel also shows the individual particle densities in the mixed
phase.}
\label{qdrop}
\end{figure}

For small droplets the quark densities are uniform to good approximation and
are determined by conditions of local chemical equilibrium. The electron
density is the same inside and outside the droplet. Thus the excess weak charge
density in the quark droplet is given by
\ber 
\rho_W &=& [(n_u-(n_d+n_s)) - (n_n-n_p)] \nonumber \\
&-& \frac{4}{3}[n_u - (n_d+n_s)+ 3 n_p] \sin^2{\theta_W} \, ,
\eer
where $ n_u,n_d, n_s$ are the up,down and strange quark number densities inside
the droplet, and $n_n,n_p$ the neutron and proton densities outside. The
individual particle densities are shown in the Fig. \ref{qdrop}. The net weak
charge and form factor for a droplet of radius $r_d=5$ fm and cell radius
$R_W=11$ fm are computed as described earlier. We find that the total weak
charge $N_W \simeq 860$ and the droplet density $N_D=18\times 10^{-5}$
fm$^{-3}$. The transport differential cross section (left panel) and the
neutrino mean free path (right panel) are shown in Fig. 5. We see that the
results are qualitatively similar to those in the $K^-$ condensate
scenario. The energy averaged neutrino mean free path $<\lambda_E>$ defined
earlier at $T=10$ MeV is reduced by a factor of $\sim 10$ compared to pure
neutron matter and at $T=30$ MeV is reduced by a factor $\sim 17$.
{\it Discussion:} There are several caveats to the study presented in this work
which we would like to address in this paragraph. First, formula for the
scattering in Eq.(\ref{formc}) treats the effects of droplet-droplet
correlations rather crudely, and short range order between droplets could
affect the low momentum transfer scattering. However, these momenta have small
weight at temperatures of interest, and corrections to the transport would be
small. Our treatment of the form factor at small $q$ may be viewed as being
conservative.  Second, in the mixed phase we studied only droplets and ignored
the possibility of other spatial structures which might exist for some range of
density. Qualitatively, the heterogeneity of the mixed phase is the key feature
that results in an enhanced neutrino opacity. Thus, our results provide a rough
estimate of what one may expect even when the structures are more complex. Both
these issues clearly require more attention and we hope to pursue this in
detail in future work. Thirdly, the droplets are likely to have low lying
collective modes which the neutrinos could excite. This will also make an
important contribution to the opacity of the material and requires further
work. Lastly, we expect that droplet formation will be fairly rapid due to the
large temperatures ($T\sim 50$ MeV) reached during the implosion of the inner
core.

Our main focus was to study the consequences of a first order phase transition
on the transport of neutrinos in the inner core of neutron stars. We have shown
that the droplet phase is indeed very opaque to neutrinos. Studies of the early
evolution of neutron stars, where diffusive transport of neutrinos directly
affects the temporal evolution of neutrino emission, which incorporate the
effects arising due to a first order phase transition remain largely
unexplored. In view of our findings in this article we may expect significant
changes to the neutrino emission if a droplet phase were to exist in the inner
core of a newly born neutron star. It seems likely that detailed studies of the
diffusive transport of neutrinos in models with a droplet phase will provide
the link between the observable features of a supernova neutrino signal and the
phases of matter at high density.

The authors would like to thank Paulo Bedaque, Chuck Horowitz, David Kaplan,
Thomas Papenbrock, Martin Savage and Ray Sawyer for useful discussions. We
would also like to thank Jason Cooke for providing us numerical tables of his
work on quark-hadron phase transitions. This work is supported by the grant
DOE-FG-06-90ER40561.

\end{document}